\documentclass[12pt]{iopart}

\usepackage{graphicx}
\usepackage{epstopdf}
\usepackage{dcolumn}
\usepackage{bm}
\usepackage{hyperref}
\usepackage{upgreek} 
\usepackage[mathlines]{lineno}

\usepackage{color,soul}

\newcommand{\bgo}{\mbox{$\mbox{Bi}_4\mbox{Ge}_{3}\mbox{O}_{12}$}}
\newcommand{\cdwo}{\mbox{$\mbox{CdWO}_{4}$}}
\newcommand{\znwo}{\mbox{$\mbox{ZnWO}_{4}$}}
\newcommand{\mmc}{\mbox{$\mbox{mm}^3$}}
\newcommand{\cms}{\mbox{$\mbox{cm}^2$}}
\newcommand{\degcelsius}{\mbox{$^\circ \mbox{C}$}}
\newcommand{\cof}{\mbox{${}^{57}\mbox{Co}$}}
\newcommand{\cso}{\mbox{${}^{137}\mbox{Cs}$}}

\newcommand{\queens}{Department of Physics, Engineering Physics \& Astronomy, Queen's University, Kingston, ON, Canada, K7L 3N6}

\newcommand{\ilm}{Univ Lyon, Universit\'e Claude Bernard Lyon 1, CNRS, Institut Lumi\`ere Mati\`ere, F-69622, VILLEURBANNE, France}

\newcommand{\insa}{Univ Lyon, INSA-Lyon, CNRS, MATEIS UMR 5510, F-69621, VILLEURBANNE, France}

\begin{document}

\title[A Multi-Channel Setup to Study Fractures in Scintillators]{A Multi-Channel Setup to Study Fractures in Scintillators}

\author{A.~Tantot${}^{1,2}$
C.~Bouard${}^1$
R.~Briche${}^1$
G.~Lef\`evre${}^{1,3}$
B.~Manier${}^1$
N.~Za\"im${}^1$
S.~Deschanel${}^3$
L.~Vanel${}^2$
P.C.F.~Di~Stefano${}^1$}
\address{${}^1$\queens}
\address{${}^2$\ilm}
\address{${}^3$\insa}
\ead{alexis.tantot@laposte.net}

\vspace{10pt}
\begin{indented}
\item[]\today
\end{indented}

\begin{abstract}
To investigate fractoluminescence in  scintillating crystals used for particle detection, we have developed a multi-channel setup built around samples of double-cleavage drilled compression (DCDC) geometry in a controllable atmosphere.  The setup allows the continuous digitization over hours of various parameters, including the applied load, and the compressive strain of the sample, as well as the acoustic emission.  Emitted visible light is recorded with nanosecond resolution, and crack propagation is monitored using infrared lighting and camera.  An example of application to \bgo\ (BGO) is provided.
\end{abstract}

\vspace{2pc}
\noindent{\it Keywords}: Brittle fracture, Fractoluminescence, Scintillators\\
\submitto{\MST}

\section{Introduction}
Fractoluminescence is the emission of light during fracture of solids. This phenomenon has been reported in brittle materials such as sucrose~\cite{Dickinson1984a}, ice~\cite{Mizuno2002}, MgO~\cite{Langford1987}, mica~\cite{Obreimoff1930}, and in soft solids such as pressure-sensitive tape~\cite{Camara2008}. 
It fits into the more general framework of fracto-emission: the emission of particles, including electrons~\cite{Langford1987}, positive ions~\cite{Dickinson1981} or X-rays~\cite{Camara2008} during fracture.

Even though fracto-emission has been known by the scientific community since Francis Bacon's experiments in 1605~\cite{Bacon1605}, it has been very little used as a tool to study the failure of materials. 
The common experimental techniques to study material failure are based on acoustic emissions or on direct observations.
Acoustic emissions allow to measure the elastic waves produced by events such as microcracking or dislocation avalanches.

In a previous article~\cite{tantot2013}, we demonstrated that in certain scintillating materials used for particle detection, fracture was accompanied by emission of light with the same spectrum as the visible scintillating light, and that fractures could therefore contribute to backgrounds in rare-event searches.  We also argued that light emission could give new insight on the rupture dynamics and provide quantitative information about the energy fracture budget. 
To fully benefit from the excellent  resolution of the photomultiplier (single photons with nanosecond timing), to be able to follow the actual crack propagation, and to control the atmosphere around the sample, we have developed an enhanced multi-channel setup.
It allows us to measure accurately light emissions, acoustic emissions, crack velocity, compression and loading force during stable and fast crack propagation in controlled atmosphere.

\section{Mechanical setup}
\subsection{Double-cleavage drilled compression geometry}\label{DCDCgeo}

Our multi-channel setup has been specially designed to study inorganic crystal scintillators such as BGO (\bgo), \cdwo~and \znwo, which are all very brittle materials. 
In order to break the specimens in a controlled and reproducible manner, the Double Cleavage Drilled Compression (DCDC) geometry defined by Janssen in 1974~\cite{janssen_specimen_1974} has been adopted. 
Each sample is cut into a $20 \times 5 \times 3~\mmc$ rectangular parallelepiped. 
All faces are polished to optical quality and a 2~mm diameter circular hole is made through the center of the $20 \times 5~\cms$ faces. 
\begin{figure} 
\centering
\includegraphics[width=0.6\linewidth]{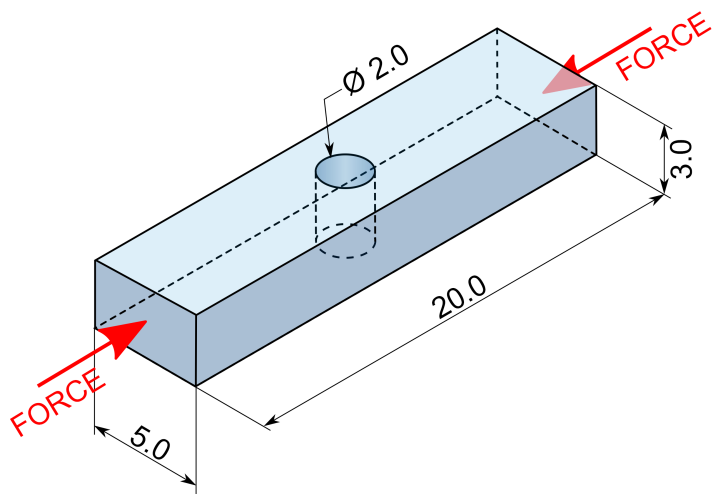}
\caption{Double-cleavage drilled compression geometry. The compression is applied along the long axis.  All dimensions are in~mm.}
\label{FigDCDC}
\end{figure}

When such a sample is compressed along its long axis (see Figure~\ref{FigDCDC}), a crack can be initiated on each side of the hole in the mirror plane parallel to the $20 \times 3~\cms$ faces. Two cracks can then propagate in opposite directions in this plane. The applied compression stress allows to control the velocity of both cracks up to a critical length, at which point the specimen abruptly breaks into two parts (fast fracture)~\cite{Pallares2009}. The slow crack propagation below the critical length is not only determined by the applied stress but also by subcritical rupture processes --- i.e. a thermally activated energy path at the crack tip which ease bond breaking. The surrounding gas is a well-known agent that may reduce the critical energy barrier required for thermally activated fracture~\cite{Rice1978}.

In order to take advantage of the sample geometry and cover the different regimes of fracture, we developed a setup which precisely controls the applied stress on the specimen and the atmosphere surrounding it.

\subsection{Setup overview}

\begin{figure}
\centering
\includegraphics[width=0.6\linewidth]{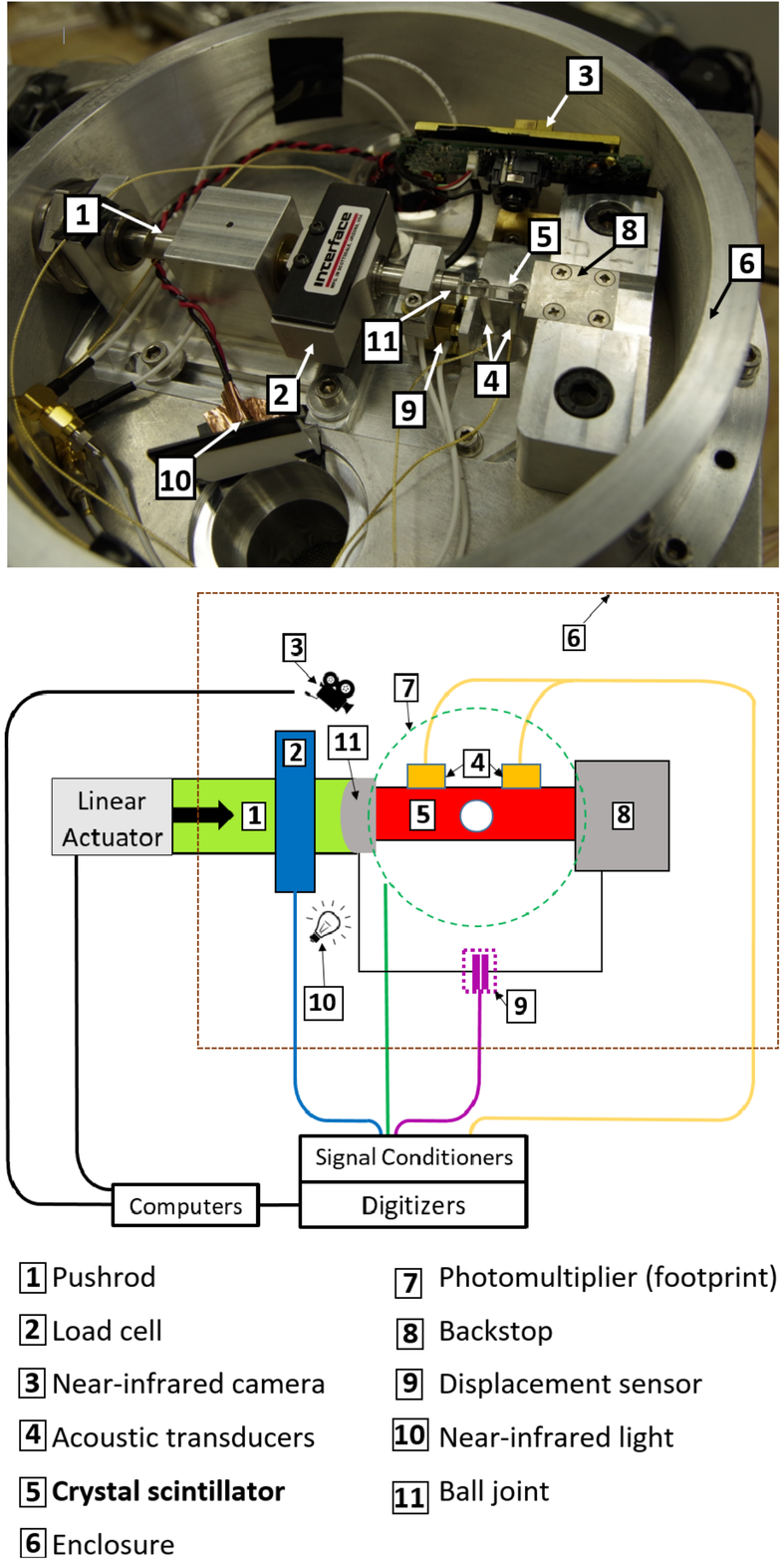}
\caption{Overview of the multi-channel setup used to compress DCDC samples.}
\label{FigSetupOverview}
\end{figure}

The setup is represented in Figure~\ref{FigSetupOverview}. 
The sample to be studied is squeezed between a backstop and a pushrod. 
The backstop is fixed and the pushrod is actuated by a linear stepper motor. 
A load cell monitors the force applied on the sample by the motor. 
The deformation of the sample is measured by a custom-made displacement sensor. 
Two acoustic sensors record the elastic vibrations produced by the crack propagation, and a photomultiplier above the sample measures the visible light emitted. Simultaneously, the propagation of cracks in the sample is observed using a near-infrared camera and a near-infrared light source (filtered and not visible by the photomultiplier). 
Signals produced by the sensors are digitized and recorded by a computer; the video is separately saved on another computer.

In order to allow accurate light measurement and to control the atmosphere, the sample, and most of the instruments, are located in a sealed chamber.  This enclosure is light-tight, with the exception of a quartz viewport for the photomultiplier tube,  and  has been tested at vacuum pressures down to $10^{-5}~\mbox{mbar}$. The vacuum is achieved using a turbo vacuum pump: HiCube~80~Eco manufactured by Pfeiffer Vacuum. The pressure is measured by a PKR~251 Compact FullRange\texttrademark~gauge.

All sensors but the photomultiplier are inside the enclosure, and were tested for vacuum operations.
The photomultiplier is outside mostly for size and cabling reasons.
The linear actuator has also been left outside because its heat could not be properly dissipated in vacuum and the grease trapped inside may not have been suitable for vacuum operations.

\subsection{Compression system}
\label{sec:CompressionSystem}

During fracture tests, the sample is squeezed between the pushrod and the backstop.
The pushrod and the backstop are actually the parts which transfer the force to the specimen. They are designed to compensate for parallel defects of crystal faces with a ball joint adapter at the tip of the pushrod.
Otherwise, the load will not properly spread on the whole faces and the edges of the specimen will potentially break.
Since the crystals we use have a relatively high Young's modulus, the pushrod and the backstop are made out of stainless steel, and the faces in contact with the specimen are mirror-polished for a good mechanical contact.

The linear stepper motor which drives the applied stress on the sample is the NA23C60 model manufactured by Zaber.
Its best step resolution (its smallest displacement) is below $0.1~\upmu\mbox{m}$ and it allows a maximum continuous thrust of 950~N.
The motor is position-controlled but its position is not relevant to determine the sample deformation.
Indeed, it would be biased by the deformation of the materials surrounding the sample.
Moreover, tests have shown some skipping
of the motor gears and they do not seem to be taken into account by the motor controller.

The applied force on the sample is regulated in real-time via an action on the stepper motor position.
The load cell is the key component which gives the feedback on the applied stress.
It converts the applied force (correlated to the stress in the double-cleavage drilled compression geometry) into an electrical signal.
The load cell is the SML-300 model with a SGA signal conditioner, both produced by Interface. Its gain is set to 7.7~mV/N for our application.

According to the manufacturer, the SML-300 load cell had never been tested at low pressure.
However, its design does not allow any air to be trapped inside the gauge, so it is  safe for tests in vacuum. 
In vacuum, the strain gauges glued on the load cell produce of the order of 1~W of heat, which can only be dissipated by conduction through the metal parts holding the gauge, and not by conduction and convection in the air.
The temperature of the cell is then expected to be slightly higher than at room pressure, but it shouldn't introduce a large error as the temperature is expected to remain within the compensated range (the thermal error is $\pm 0.0015~\%/\degcelsius$ for temperatures from $-15$~\degcelsius~to $+65$~\degcelsius).

\subsection{Displacement sensor}

A displacement sensor allows to measure the deformation of the sample, and then to get an evaluation of the work supplied by the motor.
We chose a capacitive sensor for its good resolution, and for its ability to work in vacuum and at low temperatures.
Laser-based systems were disregarded because of potential interference with the optical readout.
A capacitive sensor is made of two electrodes separated by a variable gap.  The target electrode is fixed, while the probe electrode  is attached to the point whose position is to be measured. Neglecting fringe effects, two parallel plates form a capacitance inversely proportional to their separation, $d$ the gap:
\begin{equation} \label{eq:capacitance_parallel_plate}
C_x = \frac{\epsilon_r \epsilon_0 A}{d}
\end{equation} 
where $\epsilon_r$ is the dielectric constant of the medium in the gap, $\epsilon_0 = 8.85*10^{-12}~\mbox{F}.\mbox{m}^{-1}$ the vacuum permittivity, and $A$ the surface area of each electrode. 
The dielectric constant of air is equal to $1.0006$  at normal pressure and temperature~\cite{Hector1936}, meaning that measurements in air will be consistent with those in vacuum to $0.06\%$.

\begin{figure} 
\centering
\includegraphics[width=0.6\linewidth]{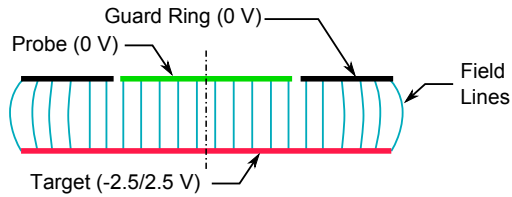}
\caption{Electrodes and field lines of the capacitive distance sensor.}
\label{FigCapaSensor_electrodeprincipe}
\end{figure}

\begin{figure}
\centering
\includegraphics[width=0.6\linewidth]{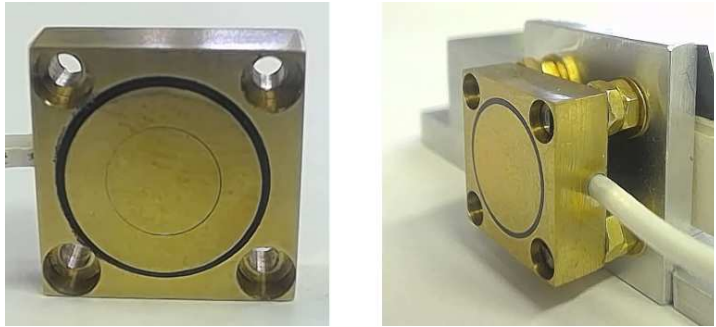}
\caption{The brass electrodes forming the capacitive distance sensor.  The probe and guard electrodes are circular and concentric, and fit into a square outer frame with screw holes for mounting.}
\label{FigCapaSensor_electrodes}
\end{figure}

Our implementation is based on the Toth-Meijer design~\cite{Toth1992}.  As shown in Figure~\ref{FigCapaSensor_electrodeprincipe}, the use of a large, concentric, guard electrode around the  small, circular, probe electrode allows to mitigate fringe effects with the larger target electrode~\cite{Toth1992,Heerens2000}.
The guard and probe electrode are isolated electrically but at the same virtual voltage.
The electric field lines are then linear between the probe and the  target electrode making the capacitance 
relatively close to the parallel plate model of Equation~\ref{eq:capacitance_parallel_plate}.
The electrodes (shown in Figure \ref{FigCapaSensor_electrodes}) have been made out of brass and are rotationally symmetrical.
The outer diameter of the target and the guard ring is 10~mm and the diameter of the probe is 6~mm.
The space between the probe and the guard ring is about $50~\upmu\mbox{m}$ and is filled with Stycast 2850~FT epoxy (Catalyst 9).
Coaxial cables connect the electrodes to the capacitance-to-period converter.

Readout is accomplished by a Martin voltage-controlled oscillator~\cite{Martin1981} that converts the capacitance into a period.
The components used for the capacitance-to-period converter are listed in Table~\ref{DispSensor_companents}. 
\newcolumntype{C}[1]{>{\centering\arraybackslash }b{#1}}
\begin{table}
\center
\renewcommand{\arraystretch}{2}
	\begin{tabular}{|C{3cm}||C{3cm}|}
	\hline
	$\mbox{C}_{\mbox{off}} = 2.7~\mbox{pF}$ & $\mbox{R} = 1~\mbox{M}\Omega$\\
	\hline	
	$\mbox{C}_{\mbox{ref}} = 4.7~\mbox{pF}$ & OA1: LTC6244\\
	\hline	
	$\mbox{C}_{\mbox{f}} = 18~\mbox{pF}$ & OA2: OPA132U\\
	\hline	
	\end{tabular}
	\caption{\label{DispSensor_companents} Components used for the modified Martin oscillator converting the capacitance into a period~\cite{Toth1992,Martin1981}.  Notations are from Ref.~\cite{Toth1992}.}
\end{table}
This board runs a Xilinx Spartan-3E FPGA which provides a basic package for fast prototyping. A $48~\mbox{MHz}$ oscillator generates the main clock for the FPGA. 
A daughter board has also been designed to connect the capacitance-to-period converter inputs and outputs to the FPGA and to carry the digital-to-analog converter (DAC).
The system output is thus a voltage proportional to the gap $d$ between the electrodes.

\begin{figure}
\centering
\includegraphics[width=1\linewidth]{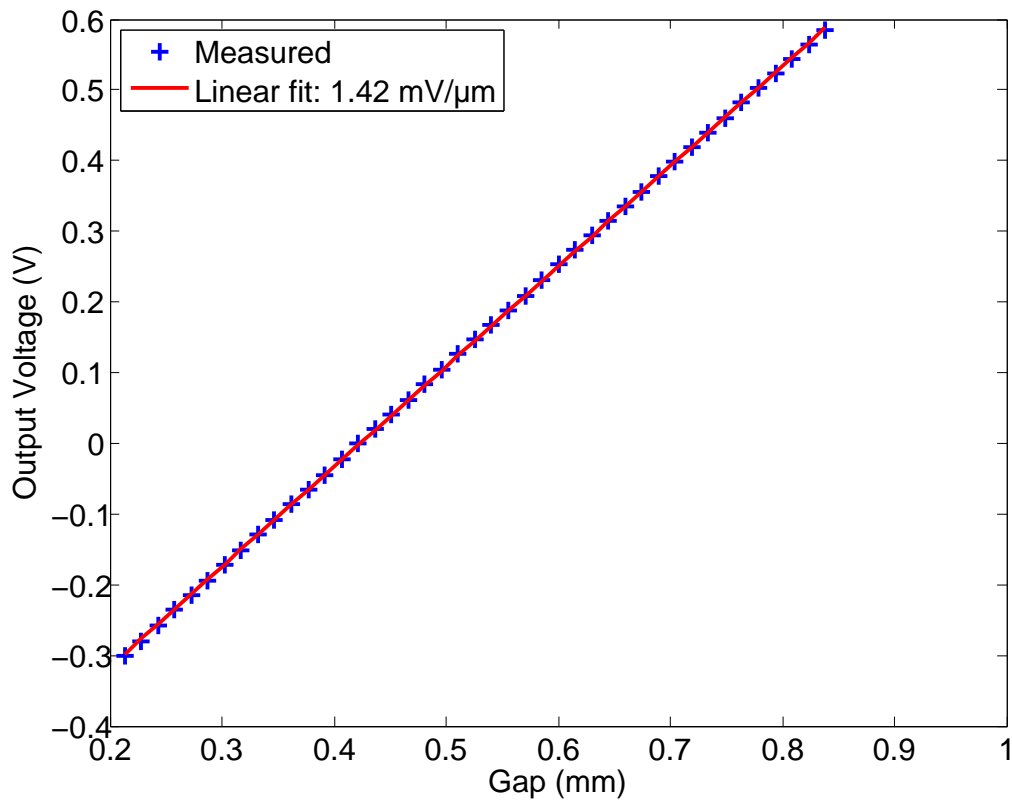}
\caption{Calibration of the displacement sensor using the linear stepper, showing the output voltage as a function of the gap between electrodes. Simulations provide the offset to convert displacement to absolute gap depicted as abscissae of both plots.}
\label{FigCapaSensor_calibration_maintext}
\end{figure}

Our capacitive sensor was calibrated using the linear stepper (Sec.~\ref{sec:CompressionSystem}) to move the probe electrode.  Results are shown in Figure~\ref{FigCapaSensor_calibration_maintext}. 
The sensor does not provide the absolute distance between the electrodes, but rather the change in distance between them as the stepper moves, which is the relevant quantity as the sample is compressed.
For the calibrations, the linear stepper is directly attached to the probe electrode (in standard operations, deformation of various intermediate objects like the force gauge and the pushrod mean that the displacement seen by the probe is different from that generated by the motor).  The linear stepper therefore provides a calibrated displacement. 
COMSOL simulations have been used to determine the actual capacitance, taking into account the full geometry of the electrodes.  Matlab was used to simulate the capacitance-to-period electronic conversion.  Together, these simulations yield the offset to convert the motor steps into the absolute distance depicted in Fig.~\ref{FigCapaSensor_calibration_maintext}.
The sensor can measure displacements, rather than the absolute gap,  with a relative accuracy of 1\%, as determined by COMSOL and Matlab simulations accounting for factors including nonlinearities and misalignment of the electrodes.  The sensor functions over gaps ranging from 0.2~mm to 0.8~mm. 
The lower end of the  range corresponds to the electrodes nearly touching;
the upper end corresponds to  the DAC saturating in amplitude. Use of updated components (Tab.~\ref{DispSensor_companents}) allows to refresh the measurement at a rate of 126~Hz,   compared to the 10~Hz of the original design~\cite{Toth1992}.
Operation has been demonstrated from ambient pressure down to  $10^{-5}~\mbox{mbar}$.

\subsection{Acoustic transducers}
\label{sec:AcoustTrans}
During material failure, crack growth and avalanches of dislocations release the stored strain energy. 
A part of this energy is converted into elastic waves, which propagates through the material. 
These elastic waves are commonly called acoustic emission 
and are generally in the ultrasonic range. 
In order to monitor the acoustic activity,  two piezoelectric transducers, converting the transient waves into an electrical signal, are placed on the surface of the specimen.

\begin{figure}
\centering
\includegraphics[width=1\linewidth]{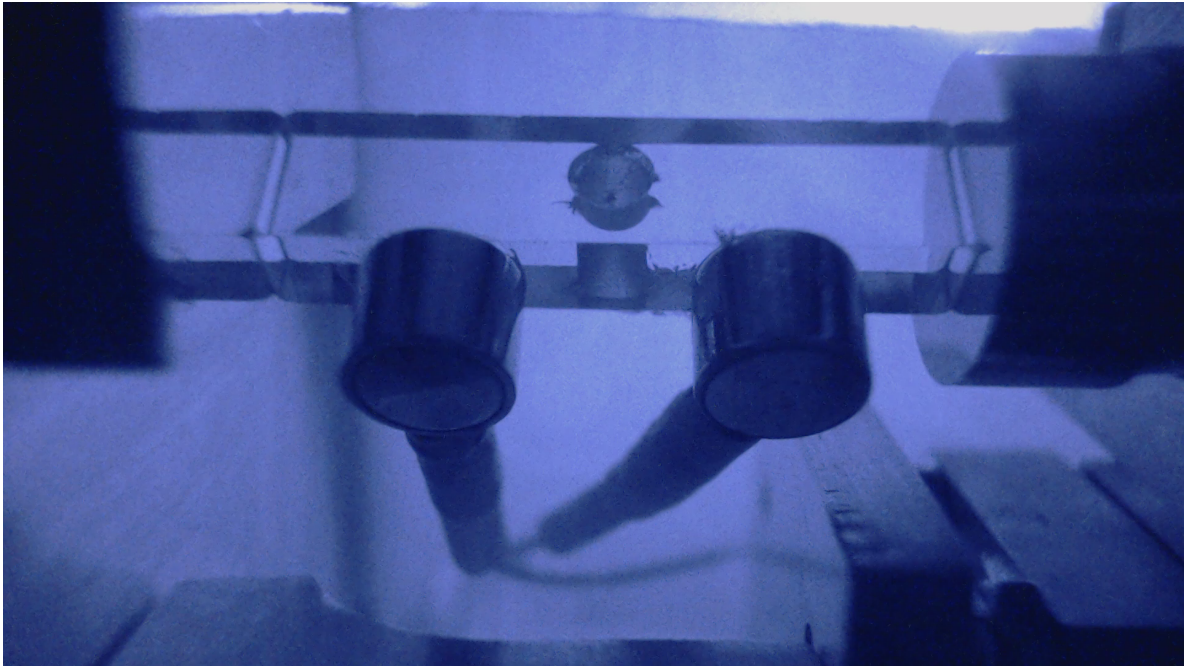}
\caption{PICO acoustic emission sensors glued on the sample.}
\label{FigAE_PICO}
\end{figure}

The PICO model, manufactured by Mistras Group and shown in Figure~\ref{FigAE_PICO}, is employed. 
It is a miniature and lightweight sensor (outside diameter 5~mm, height 4~mm, weight $<1~\mbox{g}$). 
It is well-suited for small specimens and it allows to observe a large range of frequencies, with a 20~dB bandwidth between  200~kHz and 1~MHz. 
The current generated by the acoustic sensors is converted into a voltage and amplified using a custom circuit. 
An external board cuts off all the frequencies outside the range 10~kHz--1~MHz before recording the signals.

In order to be sensitive to ultrasonic waves, the sensors require a coupling to the specimen which minimizes the density of air bubbles trapped between the specimen and the sensors~\cite{Grosse2008}.
We therefore glue the sensors to the sample using  Dow Corning~732 silicone adhesive, which offers good adhesion on multiple substrates. A small force is applied to keep the sensors in place for at least 12 hours, allowing the glue to dry. During fracture, 
no added mechanical device is needed to couple the sensors to the specimen, leaving a clear field of vision for the light sensor and for the camera.
Once a fracture test is done, the acoustic sensors can be simply removed by twisting.
The PICO sensors, adhered with the Dow Corning~732 adhesive, have been  operated successfully at  pressures down to $10^{-5}~\mbox{mbar}$.

\subsection{Photon measurements}
The light emissions during our fracture experiments are measured by a photomultiplier viewing the sample inside the experimental chamber through a quartz window. Photomultipliers use the photoelectric effect to convert an incident photon into an electron that is then multiplied into a measurable signal~\cite{Knoll2000}.
During scintillation and fracture of the BGO crystals we often use, the light emissions  are within the range 350--650~nm with a peak at 480~nm~\cite{tantot2013} (see also Sec.~\ref{Sec:Spectro}).
The R7056 model from Hamamatsu is employed because it is sensitive over the range 185--650~nm and is well-suited for photon counting.
Its maximum quantum efficiency is $20~\%$ at 420~nm.
One contribution to background comes from dark counts caused by the emission of electrons from the photocathode by thermal fluctuations (called thermionic emission). 
Since our setup does not allow to control the temperature of the photomultiplier, fracture tests are usually performed in less than two hours to avoid large fluctuations of temperature and ensuing changes in backround.

The amount of light emitted during fracture tests turned out to be quite large~\cite{tantot2013}.
In order to deal with large amounts of light, two spacers are placed in front of the photomultiplier. 
Filters such as neutral density filters or hot filters (see Section~\ref{sec:camera_tracking}) can then reduce the light reaching the light sensor. 
Between calibrations and fracture measurements, filters can be switched without changing the distance between the sensors and the specimen to be studied, in other words, without changing the solid angle of the specimen for the light sensor.

\subsection{Video tracking}
\label{sec:camera_tracking}

One of the novelties of our setup is the ability to observe the propagation of cracks within the sample, while measuring the light emission.
Based on the fact that the photomultiplier is relatively insensitive to near-infrared light, the specimen is illuminated using a near-infrared light source to allow a modified webcam to record the material failure. 
A 980~nm laser-diode (VCSEL-980 from Thorlabs) is used as near-infrared light source because it can provide enough power (2~mW) to illuminate the fracture scene and its narrow spectral emission range (Full Width at Half Maximum 0.75~nm) is outside the photomultiplier spectral range.
The narrow beam of the laser diode is spread thanks to some light-diffusing plastic sheets scavenged from a computer screen.

A modified C920 webcam from Logitech is employed to record high-definition videos \footnote{1080p (1920 pixels $\times$ 1080 pixels) with H.264 hardware encryption.} of samples during fracture tests. 
Its hot filter has been removed to broaden the camera range up to the actual limit of the CMOS sensor (above 1000~nm) and to be then sensitive to the light at 980~nm. The bulky microphones on the camera board and the blue LEDs have been unsoldered for practical reasons.
The webcam is placed inside the vacuum chamber, as close as possible to the sample.
The CMOS sensor quickly gets warm at ambient pressure, so a brass arm is used to make a thermal junction between the sensor and the metal baseplate of the setup.
In vacuum, it helps heat to dissipate by conduction.
Operation confirmed that the stock  webcam lens was not damaged when the air was pumped out of the setup, allaying fears the lens would fail because of air trapped inside.

First tests showed that the photomultiplier was definitely sensitive to the light emitted by the laser-diode. An hypothesis to explain this unexpected result is that multiple photon interactions with an electron might give to the electron enough energy to escape from the photocathode.
The relative high power of the laser-diode make this hypothesis plausible.
Indeed, a power of 2~mW means about $10^{16}~\mbox{photons}/\mbox{s}$ \footnote{The energy of a photon of wavelength $\lambda$ is $E = \frac{hc}{\lambda}$. For a 980~nm photon, $E_{980~\mbox{nm}} = 2.10^{-19}~\mbox{J}$. Hence, a $2~\mbox{mW}$ light source generates $10^{16}~\mbox{photons}/\mbox{s}$.} which is far from being negligible.
Short-pass filters have thus been added in front of the photomultiplier to attenuate the unwanted signal caused by the laser-diode.

\begin{figure}
\centering
\includegraphics[width=1\linewidth]{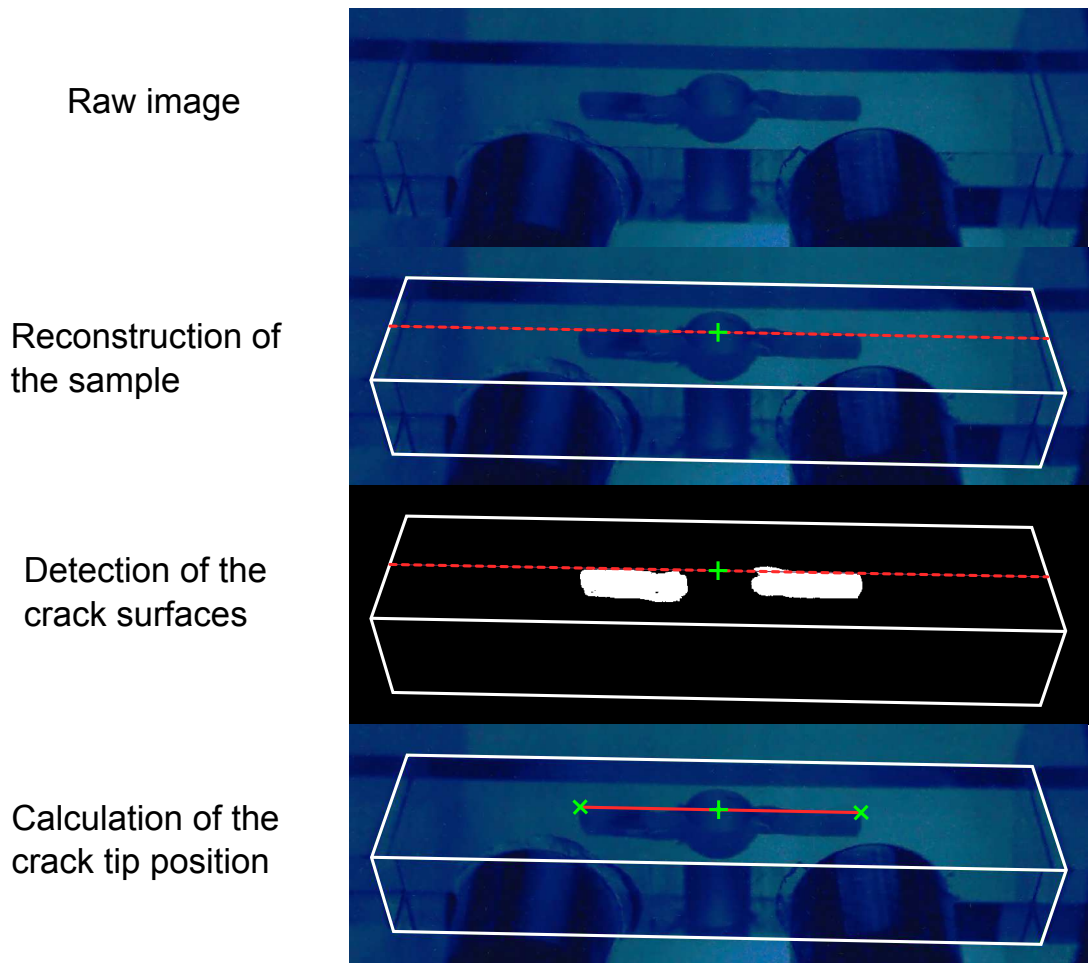}
\caption{Crack tip detection.}
\label{FigCamera_detection}
\end{figure}

During fracture tests, the sample is illuminated in such a way that the cracks are dark on a bright background (see Figure~\ref{FigCamera_detection}).
The good contrast allows to simply convert the images to grayscale and then to black-and-white using a well-chosen threshold which highlights the crack surfaces.
The crack tips are found as the horizontal extrema of the crack surfaces.
The absolute crack lengths are then computed using the pinhole camera model which describes the relationship between the 3D scene and the 2D projection on the CMOS sensor. 
This tracking algorithm has been fully implemented in Matlab.
Figure~\ref{FigCamera_detection} shows the successive steps to get a measure of the crack length.
Though pixel size corresponds to $20~\upmu\mbox{m}$ on the sample, the position of the crack front in the long dimension of the sample is known to about $\pm 0.15~\mbox{mm}$, and is dominated by the nonlinear shape of the crack front.  
In addition, our tracking algorithm assumes that the fracture strictly propagates in the longitudinal plane of the sample. However, certain cracks tend to slightly twist during their propagation, meaning the true crack length may be underestimated by up to 10\%.

During fracture tests, the frames generated by the camera have to be synchronized with the other signals (acoustic, light, force and displacement signals). 
Flashes of light are generated using the 980~nm laser-diode to mark the time on the camera.  This is done manually by closing the circuit feeding the diode by tapping it against the vacuum chamber which is the ground of the circuit.
These taps also propagate acoustic waves through the setup to the acoustic sensors.
The propagation delay is quite negligible compared to the exposure time of the camera (0.2~s) because the acoustic waves go through approximately $10~\mbox{cm}$ of aluminum at a speed of $4~\mbox{km}/\mbox{s}$ which gives a delay of only $25~\upmu \mbox{s}$. 
We have verified that any drift between the camera and the data acquisition system remains shorter than the duration of a frame on tests lasting up to 2~hours.

\subsection{Spectroscopy}
\label{Sec:Spectro}
\begin{figure}
\centering
\includegraphics[width=1\linewidth]{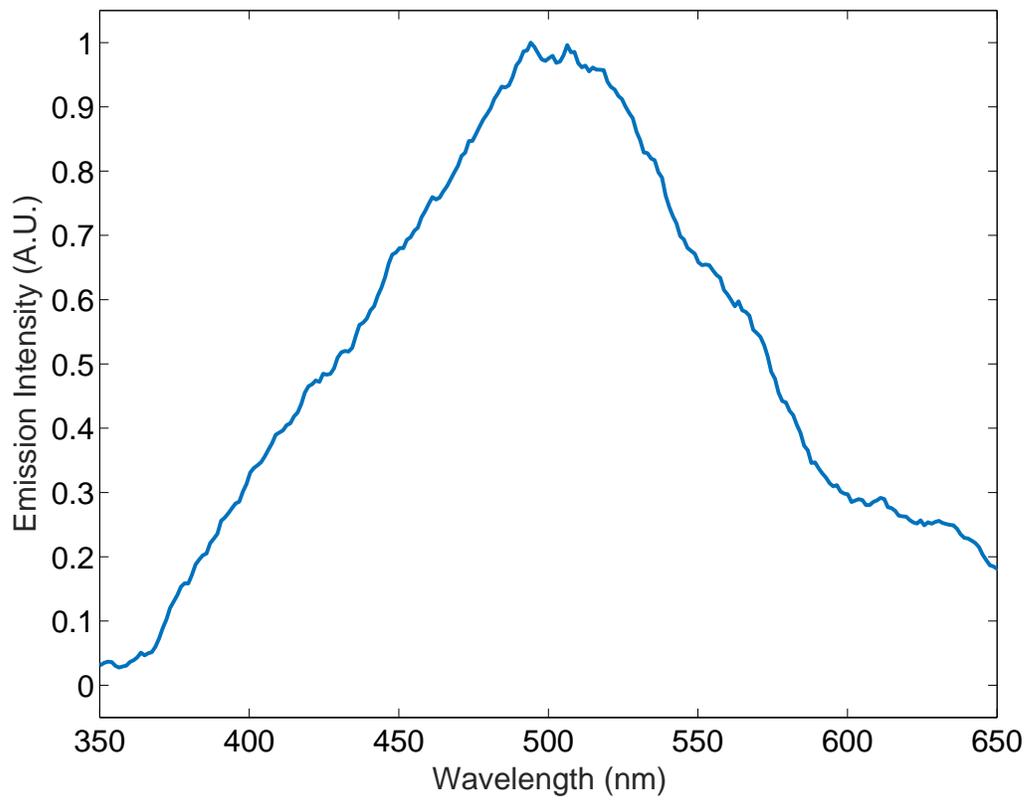}
\caption{Emission spectrum of a BGO sample fractured under a normal atmosphere at room temperature.  Wavelengths are calibrated, but no correction has been applied for efficiency of setup.}
\label{FigSpectro}
\end{figure}

In order to obtain spectroscopic information on the emitted light, at the expense of any time resolution, the setup can be optically coupled to a spectrometer.  This is done by replacing the photomultiplier by an optical fiber leading to a Horiba Micro~HR monochromator coupled to a liquid-nitrogen cooled Horiba Symphony~II CCD.  The small solid angle under which the scintillator sees the optical fiber means that light collection efficiency is much lower than with the photomultiplier. The experimental protocol is to use an exposure of a few seconds,  during which a sufficiently large force is applied to break the sample completely.  An example of a spectrum is shown in Fig.~\ref{FigSpectro}.  The spectrometer was previously  calibrated in terms of wavelengths using a mercury lamp to within 2~nm, but the spectrum amplitudes are not corrected for the various efficiencies.

\section{Data Acquisition System}

As described before, our study of fracture involves the use of multiple sensors: a photomultiplier, two acoustic sensors, a force gauge and a displacement sensor.
A data acquisition system has been developed to simultaneously digitize the signals generated by these sensors at drastically different sampling rates, and to control the force applied on the sample in real-time.
Since fracture is a one-shot process, fracture tests imply to continuously digitize all the channels, without any dead time, for hours.

\subsection{Digitizers}
Two high-speed digitizers manufactured by Cronologic~\footnote{\url{www.cronologic.de} \newline Cronologic GmbH \& Co. KG \newline Jahnstra\ss e 49 \newline 60318 Frankfurt} are used.
The Ndigo5G is a 5~giga-sample-per-second (GSPS) Analog-to-Digital Converter (ADC) with a 10 bit resolution and the Ndigo250M is a 250~mega-sample-per-second (MSPS) ADC with a 14 bit resolution.
These digitizers are designed to be plugged on PCIe \footnote{Peripheral Component Interconnect Express} slots for a high throughput.
The Ndigo5G is used to record the light signal  at full speed since single photoelectron pulses are a few ns long. The Ndigo250M is used to continuously record the slower acoustic, force and displacement signals at a reduced speed of 2~MSPS.

Two important features of these digitizers are the synchronization, done internally,  and the onboard zero suppression.
On the Ndigo5G, the onboard Field Programmable Gate Array suppresses noise between events (see Sec.~\ref{sec:LightChannel}), which significantly reduces the load on the downstream software  and hardware. In a typical fracture experiment, less than $0.1~\%$ of the light data are actually relevant, and then, automatically recorded by the data acquisition system.  Without this zero suppression, continuously sampling the light channel at 5~samples per nanosecond over an hour, would by itself lead to over 20~Tbytes of data, straining data storage and slowing offline data reduction.

\begin{figure}
\centering
\includegraphics[width=1\linewidth]{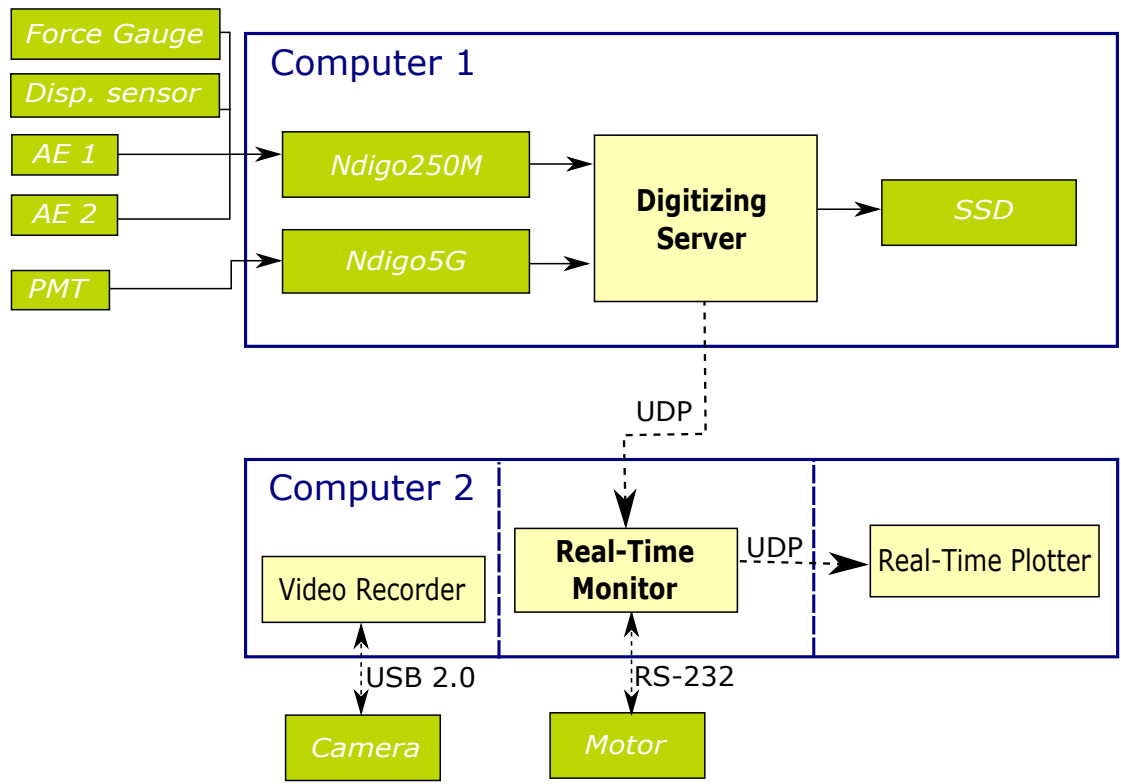}
\caption{Data acquisition system.  See text for details.}
\label{FigDigitizers}
\end{figure}

The architecture of the data acquisition system is shown in Figure~\ref{FigDigitizers}.  Different software tools have been developed in C++ using Boost and Qt. They allow to record the data generated by the digitizers, display some signals in real-time, and regulate the force applied on the specimen.
Two computers are needed to perform a fracture test.

The Ndigo5G and the Ndigo250M are plugged into Computer~1.  
The digitizing server on Computer~1 collects all the data generated by both digitizers and saves them in binary files on two solid-state drives (Samsung SSD Pro Series).
The Ndigo250M is configured to create only 4~KiB\footnote{4~kibibytes is equal to 4096~bytes.} packets of data.
This size of packets allows a fast treatment by the drives. 
The size of data packets for the light channel strongly depends on the light signal dynamic, but light data are saved only when a sufficient amount of data is collected to minimize the number of calls to the drives. 
Large buffers in the digitizer server are set in order to absorb punctual data floods caused by strong light emissions, and then spread over time the load on the solid-state drives.
 
The data from the Ndigo250M are also sent to Computer~2 over Ethernet using the User Datagram Protocol (UDP).  It is a communication protocol designed to be minimalist, it does not guarantee that data will be delivered because packets are sent without prior communications, but it reduces the latency compared to connection-oriented communication protocols.
Another advantage  is that it makes Computer~2 fully independent of Computer~1.

Computer~2 collects the data from Ndigo250M received over Ethernet. 
The force signal is extracted to regulate the force applied on the specimen using a proportional-integral-derivative controller which commands the motor setpoint. 
Acoustic, force and displacement signals are sent to MATLAB where they are displayed in order for the user to monitor fracture tests in real-time.

\subsection{Treatment of the light channel}
\label{sec:LightChannel}

\begin{figure} 
\centering
\includegraphics[width=1\linewidth]{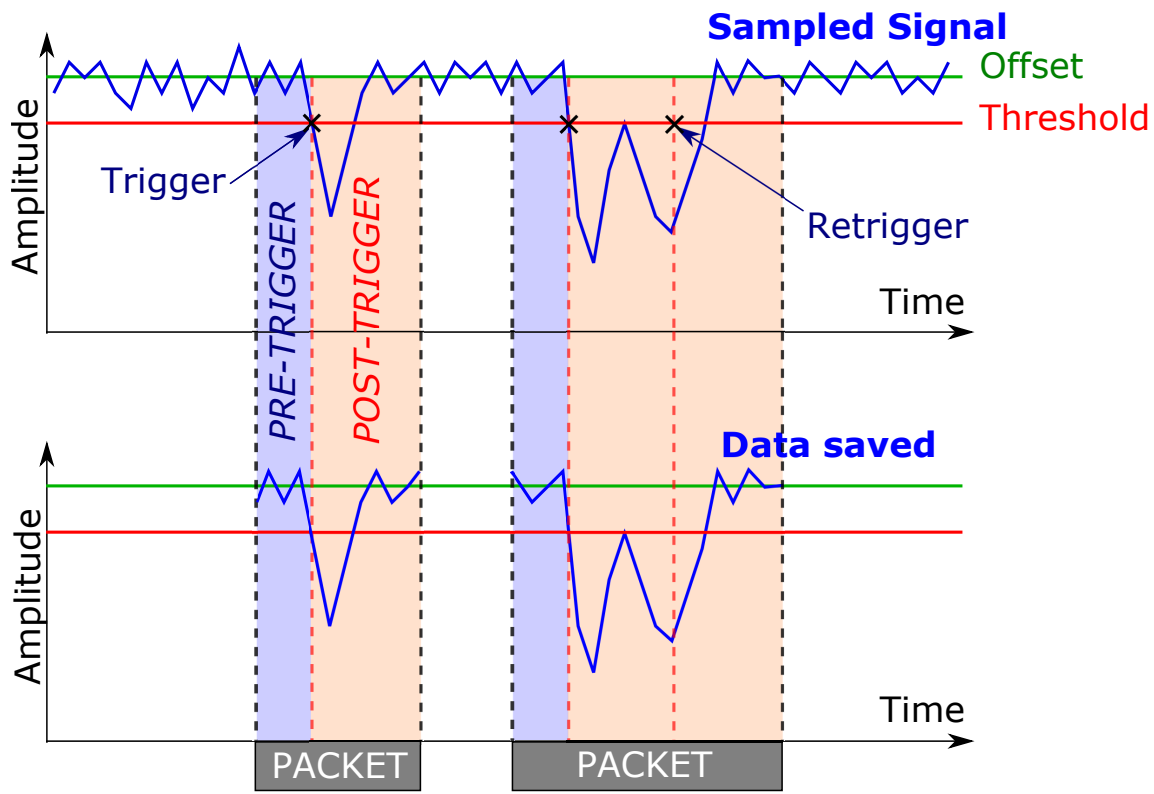}
\caption{Onboard zero suppression. When the signal goes below a given threshold, samples are saved in a structure called packet.}
\label{FigTriggering}
\end{figure}

The principle of the light acquisition is shown in Figure~\ref{FigTriggering}. 
Each photon detected by the photomultiplier produces a negative pulse on its output.
An offset is added to the signal to use the full range of the digitizer.
Onboard zero suppression allows to only record the photons.
When the signal goes below a threshold, a trigger occurs on the digitizer. For each trigger, a predefined number of signal samples before and after the trigger (pre-trigger and post-trigger, respectively) are grouped in a data packet and saved on the drives. If the signal is still below the threshold at the end of the post-trigger, the post-trigger is lengthened and the extra data samples are added to the same packet. Packets also contains information such as timestamp, number of data samples and flags (errors or warnings during the acquisition).
On the drives, only the signal samples containing useful information are then saved thanks to this onboard zero suppression.

\begin{figure} 
\centering
\includegraphics[width=1\linewidth]{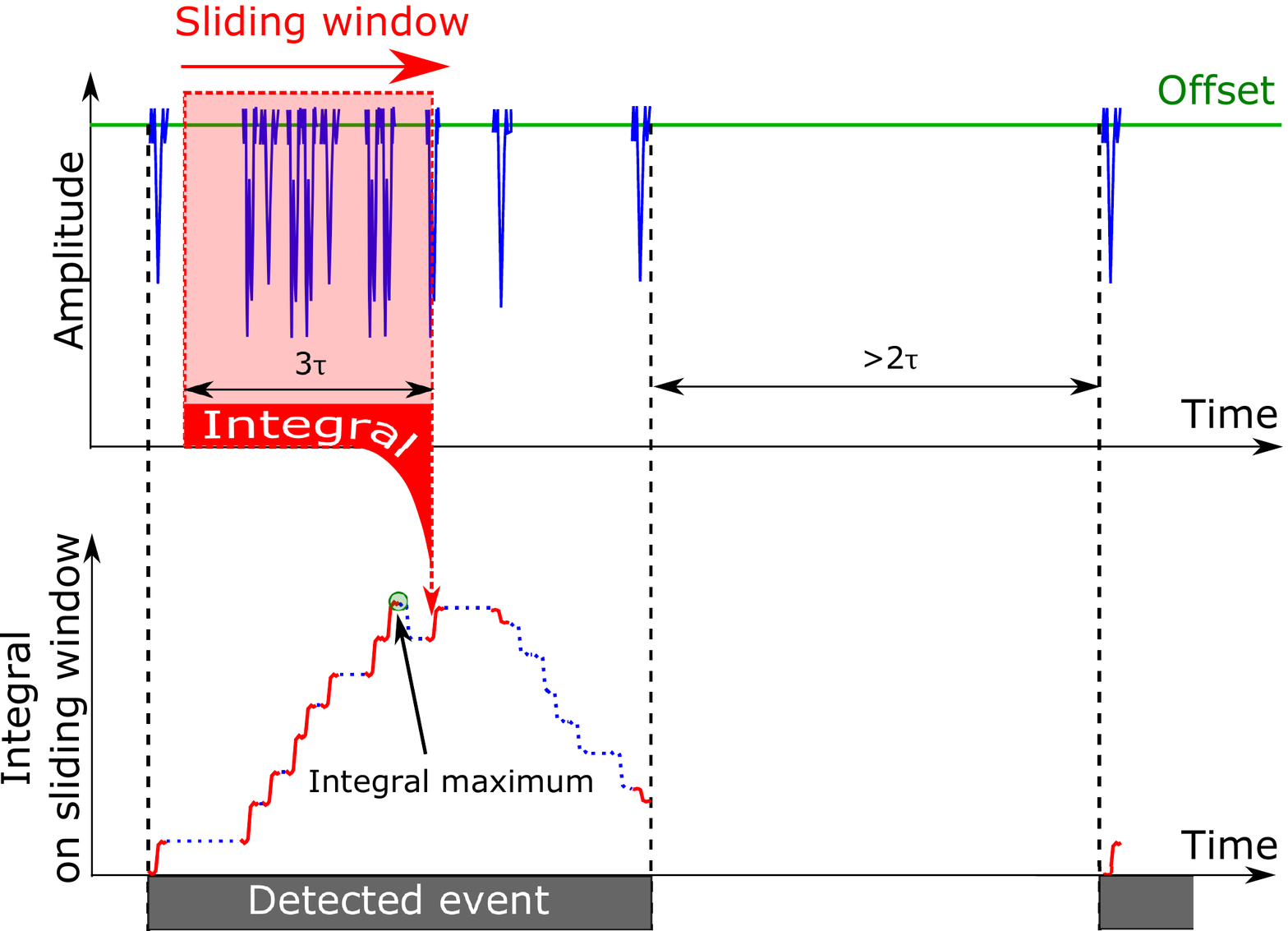}
\caption{Samples are integrated on a sliding window of $3\tau$ to obtain energy spectrum. }
\label{FigIntegrationLightEvent}
\end{figure}

When ionizing particles interact in a scintillator such as BGO, they deposit a certain amount of energy.
Some of this energy is reemitted as visible photons. 
The light yield of scintillators tells how many photons are actually emitted per unit of deposited energy.
For $\gamma$-ray sources, the light yield of BGO is 8~photons/keV~\cite{Lecoq2006}.
For a single $\gamma$ interaction, the time distribution of emitted photons generally follows one or more exponential decays.  In the case of BGO, the main time constant is $\tau = 300$~ns~\cite{Lecoq2006}.
In $3\tau$, $95~\%$ of the photons are emitted. In other scintillators, or at other temperatures, the time distribution of photons can be more complex~\cite{sivers_low-temperature_2015}. The group of photons emitted by an interaction is called a scintillation event.
To identify the scintillation events in the data, a sliding integral is computed. For each saved signal sample, all the preceding signal samples within $3\tau$ are integrated. Thus, an integral is calculated for each saved signal sample, called signal integral.
The analysis is described in Figure~\ref{FigIntegrationLightEvent}.

To identify scintillation events on the signal integral, a condition on the distance between two successive photons is used. 
If two successive photons are separated by more than $2\tau$ (called waiting time in Figure \ref{FigIntegrationLightEvent}), two different scintillation events are considered.
In the case where two scintillation events are very close to each other (successive photons separated by less than $2\tau$), a single scintillation event is considered. If this method is compared to the detection of local maxima on the integral signal (using a Schmitt trigger for example), it is not required here to set threshold(s) on the integral signal, which might depend on the photomultiplier and its voltage. 
For each scintillation event, the maximum integral is measured. It represents the maximum integral on a sliding window of $3\tau$.

\begin{figure} 
\centering
\includegraphics[width=1\linewidth]{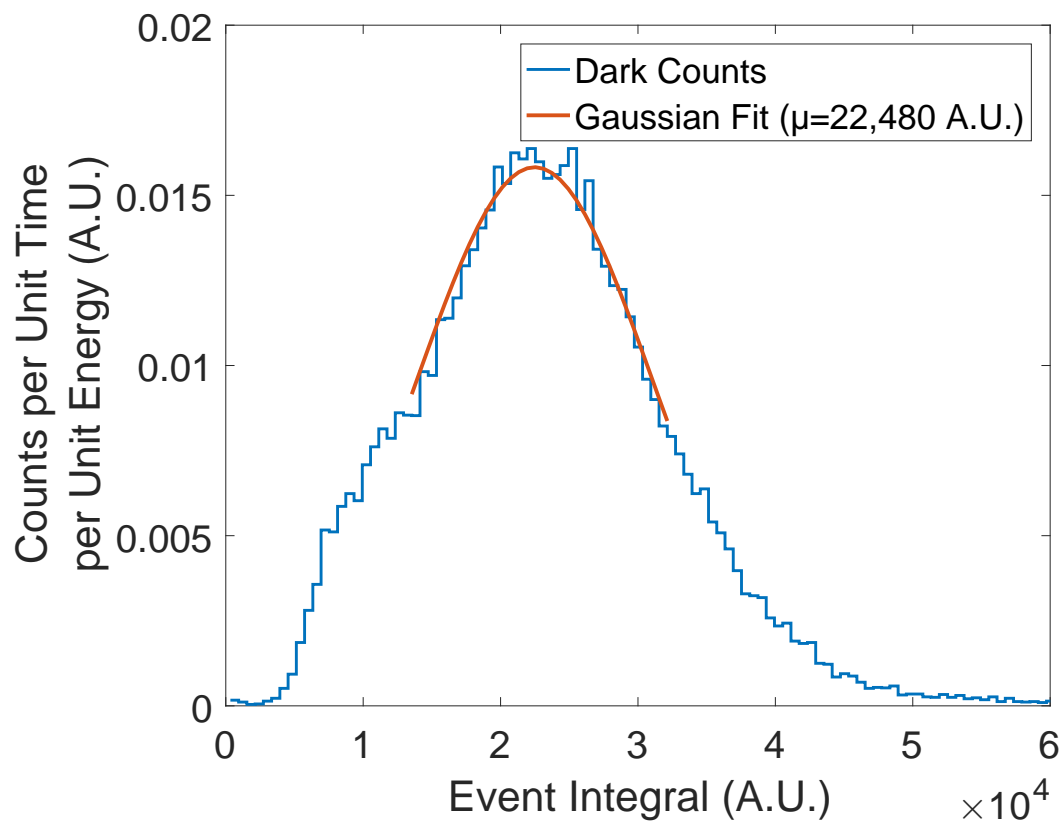}
\caption{Dark count spectrum: the photomultiplier, in a dark chamber, is excited only by thermal fluctuations. The distribution of the event integrals provides the typical value of single photo-electron integrals, taken from the central value of the Gaussian fit.}
\label{SPE}
\end{figure}

One of the first steps that needs to be carried out is determining the response of the system to single photons.
To do so, a data acquisition as described previously can be performed with the photomultiplier in a dark chamber. It is then not excited by any light source but room-temperature thermal fluctuations  eject some electrons from the photocathode. These electrons generate pulses similar to pulses generated by single incident photons.
These pulses can be easily detected as separated events by the previous analysis and their integral can be calculated.  Figure~\ref{SPE} shows the distribution of the integrals of the events from which the typical value is taken as the central value (here, 22,480 A.U.) of a Gaussian fit over part of the spectrum.
The typical rate of dark counts is a few hundred Hertz.

\begin{figure}
\centering
\includegraphics[width= 1\linewidth]{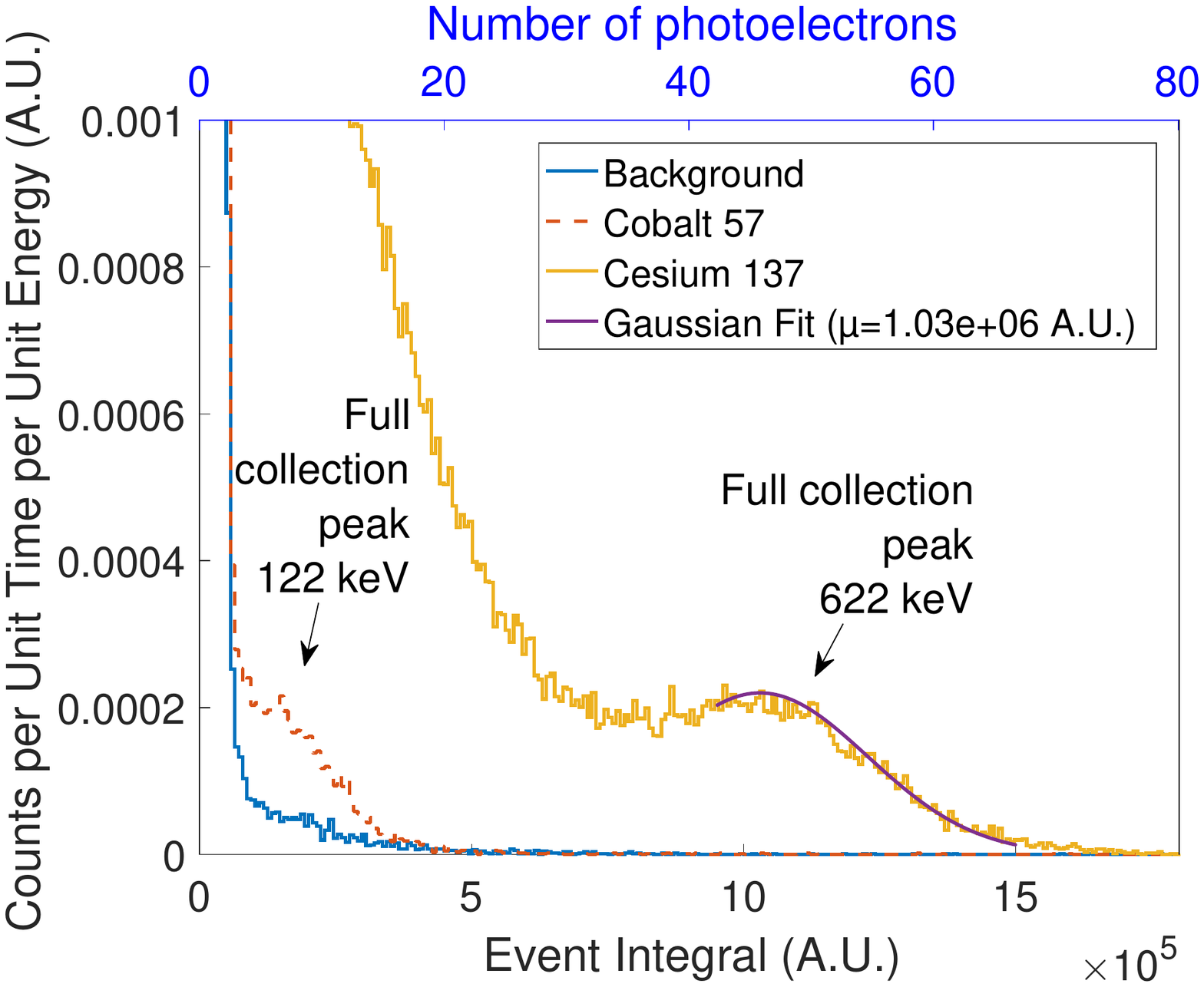}
\caption{BGO calibration spectra for various radioactive sources. Abscissae are also shown in detected photo electrons, providing an estimation of the light collection efficiency of $\sim 0.9\%$.}
\label{figCS}
\end{figure}

Next, once the scintillator is placed in the setup, and before it is fractured, the light collection efficiency of the configuration can be determined, and the light channel calibrated.
A calibration using two $\gamma$-ray sources, \cof\ (mainly 122~keV photons) and \cso\ (mainly 662~keV photons), is performed to determine a relation between the number of photons measured by the photomultiplier and the number of photons actually emitted by the sample.
The radioactive sources deposit a known energy in the sample; therefore using the light yield, one can determine the number of photons  emitted.
The light collection efficiency is determined by comparing this number to the one obtained by building the spectrum of the detected integrals, and scaling it to the response to single photelectrons.
Figure~\ref{figCS} shows typical calibration spectra that we get during an experiment. 
Thanks to the average integral of the single photoelectron obtained with Figure~\ref{SPE}, we can deduce the number of detected photoelectrons for a given event integral. Thus, for the $^{137}$Cs full collection peak of 662~keV, a Gaussian fit over part of the peak gives an average event integral of $1.03\times 10^{6}/22,480\simeq 45.8$ photoelectrons that are detected. Based on the light yield of 8 photons/keV for BGO~\cite{Knoll2000}, the light collection efficiency can be determined as $45.8/(662\times8)=0.009$. 
Since the light emission spectra during scintillation and fracture are identical~\cite{tantot2013}, the relation can be extended to the energy, so the calibration gives the ratio between the measured energy and the actual energy radiated by the specimen.

During the fracture measurement itself, the photon rate is computed.  This is achieved by integrating the light signal over a sliding window. The calculated integrals are converted into a number of emitted photons, using the light calibrations performed before the fracture test. By dividing the number of photons by the  width of the sliding window, the photon rate is thus determined. The typical width for the sliding window is about one second. This duration is a compromise, as shorter windows generally suffer from statistical fluctuations, 
 whereas longer ones do not allow to precisely observe sharp changes in the photon rate.

\subsection{Detection of acoustic events}
\begin{figure}
	\centering
    	\includegraphics[width=1\linewidth]{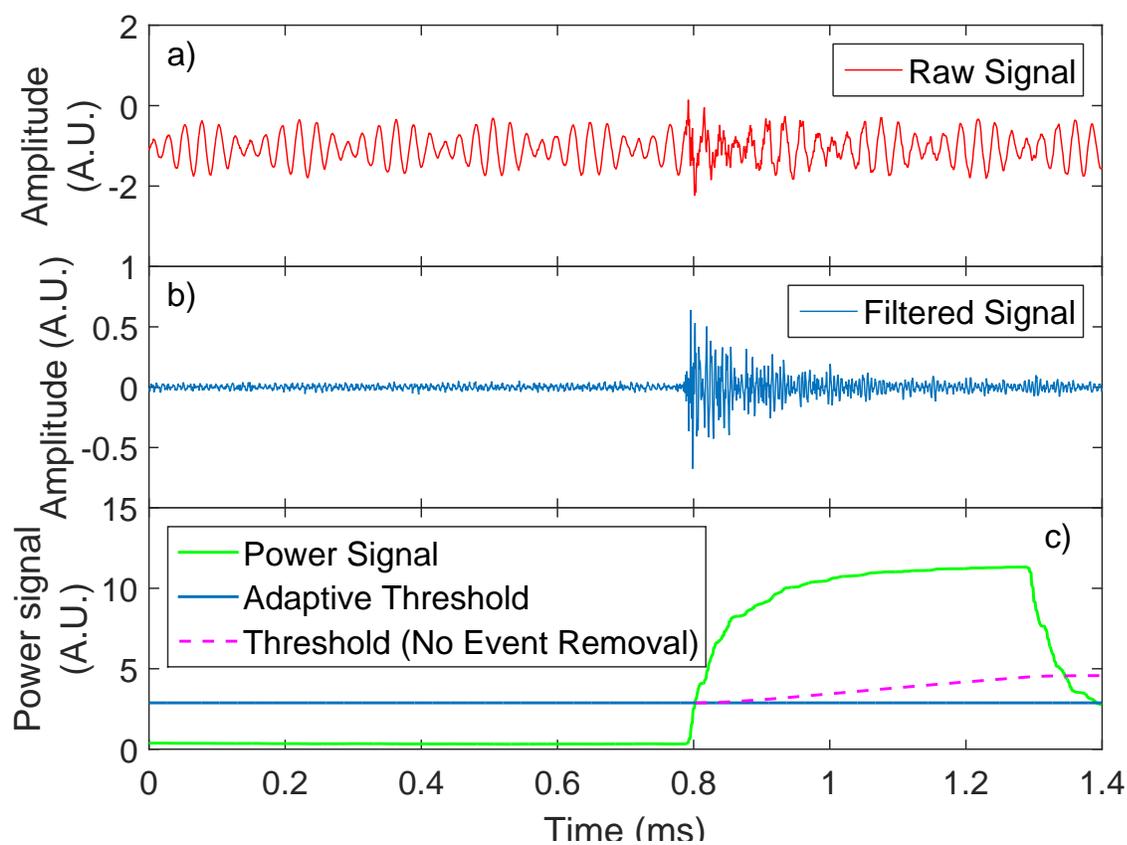}
	\caption{Detection of an acoustic event: a) raw signal recorded by the digitizer, b) digitally filtered signal (100~kHz high-pass), c) signal power (green curve) and adaptive threshold (blue curve). The acoustic event is detected when the signal power goes above the adaptive threshold. The dashed purple curve represents the adaptive threshold without the hold-on procedure during the detected event.}
	\label{AES}
\end{figure}

The two acoustic sensors (Sec.~\ref{sec:AcoustTrans}) produce analog signals which are run through a 10~kHz--1~MHz bandpass filter before being continuously recorded at 2~MSPS, as shown in Figure~\ref{AES}a.
Since the acoustic sensors are not very efficient 
below 100~kHz, a 6th-order Butterworth high-pass filter with 100~kHz cut-off frequency is used offline. It suppresses environmental noises (Figure~\ref{AES}b).

After this filtering, well-separated events are determined when the power of the signal overcomes a  threshold above the baseline  (Figure~\ref{AES}c).
The power of the data  at each instant is computed as the average of the squares of the signal samples in the previous 0.5~ms.
Next, at each point in time, the average ($m$) of the power, and its standard deviation ($\sigma$) are both computed over the previous second.  At each point in time, an adaptive threshold is then defined as $m + 15 \sigma$.  This adaptive threshold is designed to rise when the data become noisier, avoiding spurious events.  When the data surpass the threshold, the value of the threshold is frozen until the data drop again beneath it, and computation of the threshold resumes.  Individual events are defined over the interval during which the power is above the threshold. At the end of this process, all the individual events are saved to another file.

\section{Experimental protocol and illustration of operation}

Operation of the setup described above requires several steps to achieve controlled propagation of a fracture in a sample.
The sample, cut into the geometry described in Sec.~\ref{DCDCgeo}, requires extra preparation. First, the edges in contact with the backstop and the ball joint are sanded down, which avoids unwanted microcracks at the sharp edges during compression.
Then, using a surgical blade, small notches are made on either side of the hole and on each face of the sample, in the direction in which the fracture will propagate. These artificial flaws will help to initiate the cracks. Next, the sample is cleaned, and the acoustics sensors are glued as described in Sec.~\ref{sec:AcoustTrans}. 

Once the crystal is prepared, it is carefully positioned in the setup between the ball joint and the backstop. It must be precisely centered on the loading axis in order to avoid deviations of the crack from the nominal propagation plane. The specimen is also tilted so that the infrared camera can monitor the whole crack propagation.
Next, the enclosure is closed which allows to  control the gases present during the fracture test and their pressures. The sample is illuminated with infrared light and observed by the camera. 
The data acquisition system is then started, which records all the sensors: load cell, displacement sensor, acoustic sensors, photomultiplier and camera.
The sample is then ready for a controlled crack growth experiment.

\begin{figure*} 
\centering
\includegraphics[width=1\linewidth]{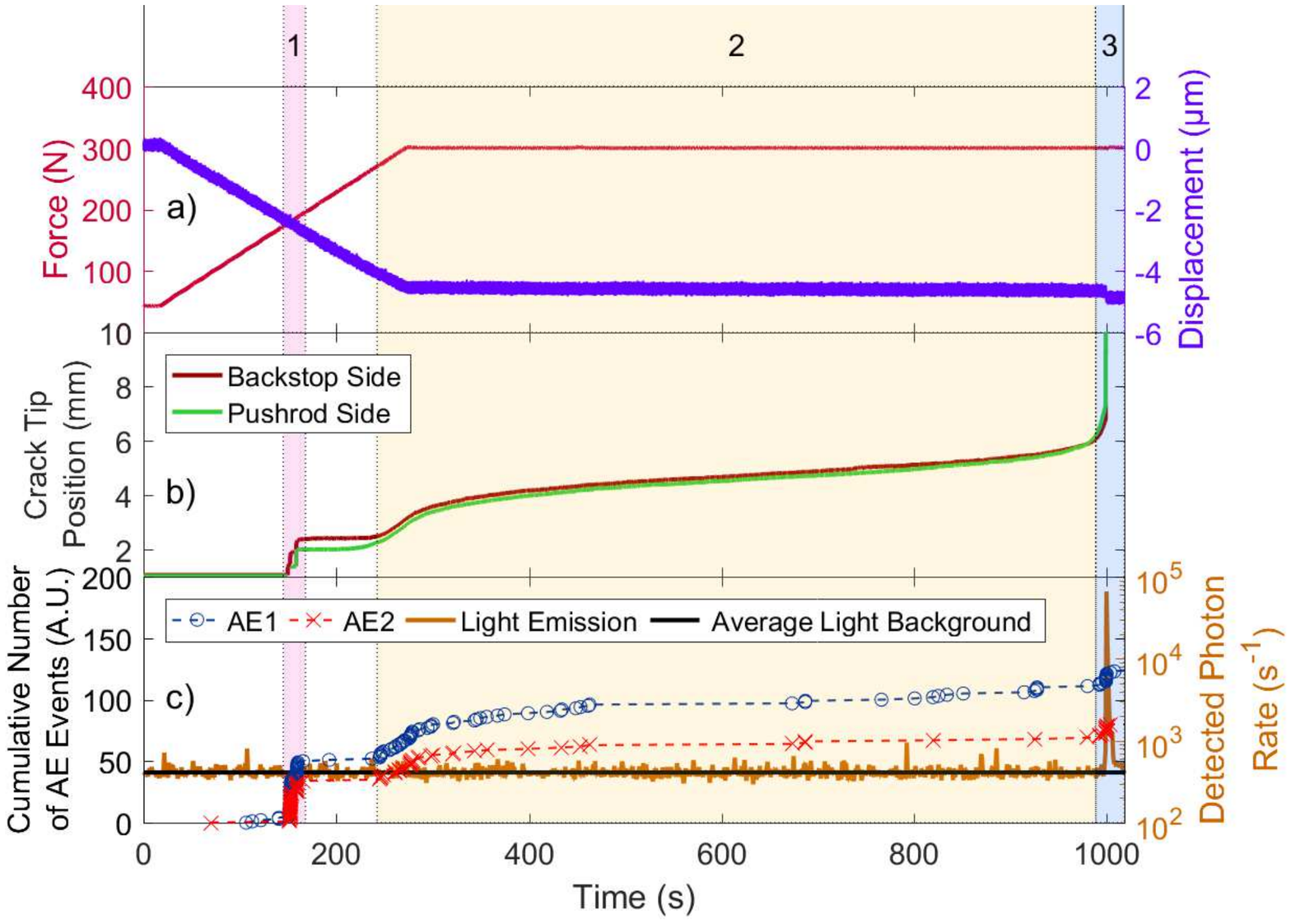}
\caption{Example of results from a BGO sample under a normal atmosphere, illustrating the various measurement channels.  Notable regimes are 1) reopening of previously initiated crack, 2) slow crack growth, 3) fast fracture.}
\label{BulleGaryPlot}
\end{figure*}

A typical fracture test at ambient pressure with a BGO sample is presented in Figure~\ref{BulleGaryPlot}. 
An initial force of $50$~N is applied.
The force is slowly increased  at $1$~N/s up to $300$~N (around $240$~s), and is then kept constant by the regulation system (Fig.~\ref{BulleGaryPlot}a, pink curve). 
The opposite of the  displacement,  measured by the capacitive sensor (Fig.~\ref{BulleGaryPlot}a, purple curve), gives the variation of sample  compression.
The compression increases  by about $5$~$\upmu \mbox{m}$ during the force ramp.
Afterwards, the displacement decreases slowly except for a jump that occurs at $1000$~s and that is associated to fast and full rupture of the sample into two halves.
The propagation of the two opposite cracks observed by the infrared camera is shown in Figure~\ref{BulleGaryPlot}b.  The center of the sample is chosen as origin, hence the crack length only starts at a distance of $1$~mm which is the edge of the central hole.
At $180$~s, a jump in crack growth is observed, corresponding to the reopening of the   crack which had been initiated previously and allowed to close. Then, while the force is kept constant, a slow growth of the crack is observed up to a critical length (near $990$~s) when it accelerates and cleaves the material in two. Light 
and acoustic emissions 
are shown in Figure~\ref{BulleGaryPlot}c. 
Acoustic emissions are correlated with the whole crack propagation. In this setup, light emissions are prominently observed at the moment of fast fracture.  These correlations confirm our previous observations~\cite{tantot2013}. Full analysis of these results, and a host of  experiments, are underway.

\section{Conclusion}

In order to study fractures in scintillating materials, we have developed a novel multi-channel setup under controlled atmosphere. Using the DCDC geometry for the sample, our setup allows the simultaneous measurement of the force applied to the sample, the compression, the acoustic and visible light emissions, all the while filming the propagation of the fracture under infrared lighting.  The data acquisition system enables continuous recording of all channels, allowing analysis of rupture dynamics over extremely large time and energy ranges.  For instance, nanosecond timing is achieved for the light channel, with onboard zero supression to limit the amount of data stored.  Various loading profiles can be applied, allowing the study of different fracture regimes. 

Thanks to this setup, we will be able to precisely explore light emision during the subcritical regime in particular.  We also plan to study effects such as that of the atmosphere on the fracture and light emission.  Lastly, our setup is not restricted to scintillators, but could be used for other materials such as PMMA and glass~\cite{Pallares2009}.

\section*{Acknowledgments}
We thank Gary Contant and Chuck Hearns (Queen's University) for contributing to the mechanical development, and Jean-Michel Combes and Fran\c cois Gay (Universit\'e de Lyon) who developed the acoustic amplifiers and participated to the electronic design. 
This work has been funded in Canada by NSERC (Grant No. SAPIN 386432), CFI-LOF and ORF-SIF (Project No. 24536), and by the France-Canada Research Fund
(Project ‘‘Listening to Scintillating Fractures’’). Alexis Tantot has been supported by an Explora'Doc grant from Rh\^ones-Alpes (France).

\end{document}